\documentstyle[12pt]{article}
\ifx\text\undefined
\let\text\textrm    %% for latex2e by xxx e-print admin 9/97 
\fi

\begin{document}

\title{Turbulent Heating in the Solar Wind and in the Solar Corona}
\author{Mahendra K. Verma  \\ %EndAName
Department of Physics, Indian Institute of Technology,\\Kanpur 208016, India}
\date{\today }
\maketitle

\begin{abstract}
In this paper we calculate the turbulent heating rates in the solar wind
using the Kolmogorov-like MHD turbulence phenomenology with Kolmogorov's
constants calculated by {\it Verma and Bhattacharjee }[1995b,c]. We find
that the turbulent heating can not account for the total heating of the
nonAlfv\'enic streams in the solar wind. We show that dissipation due to
thermal conduction is also a potential heating source. Regarding the
Alfv\'enic streams, the predicted turbulent heating rates using the
constants of {\it Verma and Bhattacharjee }[1995c] are higher than the
observed heating rates; the predicted dissipation rates are probably
overestimates because Alfv\'enic streams have not reached steady-state. We
also compare the predicted turbulent heating rates in the solar corona with
the observations; the Kolmogorov-like phenomenology predicts dissipation
rates comparable to the observed heating rates in the corona [{\it Hollweg, }%
1984], but Dobrowoly et al.'s generalized Kraichnan model yields heating
rates much less than that required.
\end{abstract}
\pagebreak

\section{Introduction}

The proton temperature $T$ of the solar wind, regarded here as a single
magnetofluid parameter, is observed to decrease slower that what adiabatic
cooling would predict [{\it Schwenn}, 1983; {\it Marsch et al., }1983; {\it %
Schwartz and Marsch}, 1983; {\it Gazis}, 1984; {\it Freeman and Lopez},
1985; {\it Lopez and Freeman}, 1986;{\it \ Freeman et al.}, 1992]. These
observations indicate that the protons in the solar wind is heated in its
transit. Note that in this paper we will not discuss the temperature
evolution of electrons and alpha particles. Also, since the solar wind has
complicated flow behaviour, the simplistic arguments presented in this brief
report are of a qualitative nature.

Various attempts have been made to explain the observed heating. The
proposed sources of heating are shocks [{\it Whang et al.}, 1990],
turbulence [{\it Tu et al., }1984; {\it Tu, }1987; {\it Tu, }1988; {\it %
Verma et al.}, 1995a], heat conduction [{\it Gazis, }1984], interactions
with neutral particles [{\it Isenberg et al.}, 1985] etc. {\it Whang et al. }%
[1990] performed observational and simulational studies on the corotating
shocks in the solar wind between 1 and 15 AU, and showed that in this
region, shocks are major heating sources in the solar wind, with most of the
heating confined to 1-5 AU. They found that the increase in the entropy$\ \,$%
per shock is approximately $0.8\times 10^{-23}$ J/K/proton. Note that
entropy is defined as $k_B\ln (T^{1.5}/n)$, where $T$ is the temperature, $n$
is the particle density, and $k_B\,$ is the Boltzmann constant. They also
derived that the temperature variation due to the observed entropy increase
yields $T(r)\propto r^{-0.53}$, where $r$ is the distance from the sun. Note
that without any heating, the temperature of the solar wind would be
proportional to $r^{-4/3}$.

In a recent paper, henceforth referred to as paper I, {\it Verma et al.}
[1995a] calculated turbulent heating in the solar wind using the existing
MHD turbulence phenomenologies, Kolmogorov-like MHD turbulence phenomenology
[{\it Marsch}, 1990; {\it Matthaeus and Zhou}, 1989; {\it Zhou and Matthaeus}%
, 1990] and Dobrowolny et al.'s generalized Kraichnan phenomenology [{\it %
Dobrowolny et al.}, 1980; {\it Kraichnan}, 1965]. The basic idea paper I is
similar to that of {\it Tu et al.} [1984] and {\it Tu }[1988] which is that
the dissipation rates can be calculated from the inertial range energy
spectra without any knowledge of the details of dissipation mechanism. There
are some differences between our model and {\it Tu'}s [1988] model; they are
compared in paper I. In paper I the constants of the phenomenologies were
treated as free parameters. Their results were in general agreement with the
observations for certain values of the free parameters. Recently {\it Verma
and Bhattacharjee} [1995b,c] have calculated the constants of the
Kolmogorov-like phenomenology theoretically using the direct interaction
approximation (DIA) technique of Kraichnan [{\it Kraichnan, }1959]. The
constants calculated theoretically are in general agreement with the results
of numerical simulations by {\it Verma }[1994] and {\it Verma et al.}
[1995d]. However, the constants calculated by DIA and numerical simulations
are different from the ones used in paper I. In this paper we calculate the
dissipation rates using the constants calculated theoretically, contrary to
the procedure of paper I in which the constants were basically free
parameters. Section 2 of this paper contains these calculations. Note that
the dissipation rates in the solar wind varies with distances. In this paper
we estimate typical dissipation rates in the solar wind (at 1 AU).

In section 3 we estimate the contributions of turbulent thermal conduction
to the heating in the solar wind. In section 4 of this paper we have
estimated turbulent dissipation rates in the solar corona using Dobrowolny
et al.'s generalized Kraichnan phenomenology. Here we also restate {\it %
Hollweg}'s [1986] arguments in terms of  the Kolmogorov-like MHD turbulence
phenomenology. Section 5 contains the concluding remarks.

\section{Turbulent Heating in the Solar Wind}

In paper I we have derived a formula for the temperature evolution of the
solar wind protons due to turbulent heating. Other source of heating were
not included in that paper. However, one can easily generalize the equation
for the temperature evolution when heating due to shocks and heat conduction
are also included [see {\it Priest, }1982]. The generalized equation is
\begin{equation}
\label{tempr}\frac{dT}{dr}+\frac{2R}{C_V}\frac Tr=\frac \Sigma {UC_V}\left(
\epsilon _{turb}+\epsilon _{shock}+\frac{\nabla \cdot {\bf q}}\rho \right) ,
\end{equation}
where $T$ is the temperature, $r$ is the radial distance from the sun, $R$
is the Rydberg's constant, $C_V$ is the specific heat per unit mass at
constant volume, $\Sigma $ is the mass per unit mole ($\approx $1 gm for the
solar wind plasma), $U$ is the solar wind mean speed, $\epsilon _{turb}$ is
the turbulent dissipation rate per unit mass, $\epsilon _{shock}$ is the
dissipation rate per unit mass due to shocks, ${\bf q}${\bf \ }is the heat
flux, and $\rho $ is the density. Here as well as in paper I we have assumed
constant speed for the solar wind, hence $\rho \propto r^{-2}$. We make this
assumption to simplify the analysis even though the density is observed to
vary somewhat differently from $r^{-2}$. The Eq. (\ref{tempr}) has the
immediate solution
\begin{equation}
\label{Tr}T(r,U)=\left( \frac r{r_0}\right) ^{-4/3}\left[ T_0(U)+\frac
\Sigma {UC_V}\int_{r_0}^r\left( \frac s{r_0}\right) ^{4/3}\left( \epsilon
_{turb}+\epsilon _{shock}+\frac{\nabla \cdot {\bf q}}\rho \right) ds\right]
\end{equation}
where $T_0(U)$ is the temperature at $r_0$ for a stream with velocity $U.$

In paper I we have calculated the turbulent heating in the solar wind using
the Kolmogorov-like MHD turbulence phenomenology and Dobrowolny et al.'s
modified Kraichnan's phenomenology. Here we state the energy spectra in
these phenomenologies. In Kolmogorov-like phenomenology, they are
\begin{equation}
\label{kolmlike}E^{\pm }(k)=C^{\pm }\left( \epsilon ^{\pm }\right)
^{4/3}\left( \epsilon ^{\mp }\right) ^{-2/3}k^{-5/3}
\end{equation}
where $E^{\pm }(k)$ are the energy spectra of fluctuations ${\bf z}^{\pm }=%
{\bf u\pm b}$ (${\bf u}$ is the velocity fluctuation, and ${\bf b}$ is the
magnetic fluctuation in velocity units), $\epsilon ^{\pm }\ $are the
dissipation rates of ${\bf z}^{\pm }$, and $C^{\pm }$ are Kolmogorov's
constants for MHD turbulence. By inverting the above equation we can obtain
the total dissipation rate which is
\begin{equation}
\label{Kolm-diss}\epsilon =\frac{\epsilon ^{+}+\epsilon ^{-}}2=\frac
12\left( \frac{\alpha _1}{C^{-}\sqrt{C^{+}}}+\frac{\sqrt{\alpha _1}}{C^{+}%
\sqrt{C^{-}}}\right) \left[ E^{+}(k)\right] ^{3/2}k^{5/2},
\end{equation}
where $\alpha _1=E^{-}/E^{+}$. Note that using this scheme we can obtain the
dissipation rates using the energy spectra of the wind and the constants $%
C^{\pm }$ without any knowledge of  the details of dissipation mechanism.

In this paper, without any loss of generality we take $\alpha _1\leq 1$. The
normalized cross helicity $\sigma _c$, a measure of velocity and magnetic
field correlation, is defined as%
$$
\sigma _c=\frac{2{\bf v\cdot b}}{v^2+b^2}=\frac{1-\alpha _1}{1+\alpha _1}.
$$
The normalized cross helicity plays an important role in determining
dissipation in the solar wind [{\it Verma et al., }1995a].

Recently {\it Verma and Bhattacharjee} [1995b,c] have calculated the values
of the constants $C^{\pm }$ using DIA{\bf . }They find that the $C^{\pm }$
are not universal constants as is the case in fluid turbulence, but they
depend on the Alfv\'en ratio $r_A$ (ratio of kinetic and magnetic energy)
and the normalized cross helicity $\sigma _c$. In Table 1 we list some of
the values of these constants. In these calculations they take $E^{+}>E^{-}$%
. Unfortunately, in paper I it was assumed that $C^{+}=C^{-}=C$, an
approximation valid only for nonAlfv\'enic streams. In this paper we will
use constants $C^{+}$ and $C^{-}$ obtained by DIA.

In Dobrowolny et al.'s generalized Kraichnan's phenomenology, the
dissipation rates are given by
\begin{equation}
\label{Dobro}\epsilon ^{+}=\epsilon ^{-}=A^{-2}B_0^{-1}E^{+}(k)E^{-}(k)k^3
\end{equation}
where $A$ is Kraichnan's constant, and $B_0$ is the mean magnetic field or
the magnetic field of the largest eddies. {\it Verma and Bhattacharjee }%
[1995c] have also calculated the Kraichnan's constant for a variation of
{\it Dobrowolny et al.}'s model and found that the Kraichnan's constant $A$
is approximately 2.0 for nonAlfv\'enic streams. In this paper we emphasize
the Kolmogorov-like phenomenology because its predictions appear most
consistent with the observed energy spectra of $-$5/3 for both ${\bf z}^{\pm
}$ in the solar wind [{\it Matthaeus and Goldstein, }1982; {\it Marsch and
Tu, }1989].

{}From the energy spectra and the constants $C^{\pm }$ we can estimate the
dissipation rate $\varepsilon _{turb}$ at reference position $r_0$ using Eq.
(\ref{Kolm-diss}). Then we can obtain the temperature at any position $r$
using Eq.~(\ref{Tr}). In paper I the best fit to the observed temperature
evolution of the solar wind was found when $C=1.0$ for nonAlfv\'enic wind
and $C=8.0$ for Alfv\'enic wind, and the dissipation rates at 1\ AU was
found to be of the order of $10^{-3\text{ }}$km$^2$/s$^3$. The calculation
of the temperature evolution using {\it Dobrowolny et al.}'s model with
neither $A=1$ nor $A=0.5$ provided a good fit to the observed temperature
evolution, however here too for $A=1$, the dissipation rates at 1 AU were of
the order of $0.5\times 10^{-3\text{ }}$km$^2/$s$^3$ [{\it Verma et al.},
1995a]. See paper I for details.

The turbulent dissipation rate depends on the constant $C$, and it is
crucial to use the correct constants $C$ for proper estimation of turbulent
heating. In the following subsections we use the constants calculated by
{\it Verma and Bhattacharjee} [1995b,c] to estimate the turbulent
dissipation rates in the solar wind.

\subsection{NonAlfv\'enic Streams}

In the outer heliosphere most of the solar wind streams have been observed
to be nonAlfv\'enic ($\sigma _c\sim 0$) with $r_A\sim 0.5$. From Table 1 $%
C=4.04$ [{\it Verma and Bhattacharjee, }1995b,c] for these streams. In paper
I {\it Verma et al.} found that when $C=1.0$, turbulent heating can account
for all the heating in the solar wind. They estimated turbulent dissipation
rate at 1 AU when $C=1$ to be approximately $10^{-3}$ km$^2$s$^{-3}$.
However, since theoretical value of $C$ is approximately 4 in the outer
heliosphere [{\it Verma and Bhattacharjee, }1995b,c], the correct turbulent
dissipation rate is approximately (refer to Eq. (\ref{Kolm-diss}))
\begin{equation}
\epsilon =\epsilon _{paperI}\left( \frac{C_{paperI}}C\right) ^{3/2}\sim
\frac{10^{-3}}{4^{3/2}}\sim 1.25\times 10^{-4}km^2s^{-3}.
\end{equation}
Hence, in the outer heliosphere, turbulent heating is not sufficient to heat
the plasma to the observed temperature. The rest of the heating can be
provided by shocks, thermal diffusion, and other mechanisms. {\it Whang et
al.} [1990] showed that the shock heating is significant, while {\it Gazis }%
[1984] showed that heating due to thermal conduction is significant (for
thermal conduction, also refer to {\it Marsch et al. }[1983]; {\it Marsch
and Richter }[1984]). We briefly sketch the results of their work in the
section 3.

In the inner heliosphere, the fluctuations are somewhat fluid dominated.
Therefore, $C$ in the inner heliosphere is smaller than $C$ of the outer
heliosphere (see Table 1), hence resulting in a larger turbulent heating. If
we take $r_A$ in the range of 1 to 2, $C$ is in the range of 3.38 to 2.58
(see Table 1). With these constants too the turbulent heating is utmost $%
C^{-3/2}\sim 25-30\%$ (approximately one-quarter) of the of observed heating
(heating rate of paper I) in the solar wind. The corotating shocks are
absent in the inner heliosphere. Therefore, shock heating is presumably not
significant for nonAlfv\'enic streams in the inner heliosphere. Hence,
neither shocks nor turbulence can provide enough heating in the inner
heliosphere. In section 3 we will estimate the contributions of thermal
conduction to the overall heating in the inner heliosphere.

Regarding {\it Dobrowolny et al.}'s model, for nonAlfv\'enic streams, the
Kraichnan's constant $A$ is approximately 2.0. With $A=2.0,$ the dissipation
rates at 1 AU will be of the order of $\epsilon _{paperI}/A^2\sim 0.5\times
10^{-3}/A^2\sim 10^{-4}$ km$^2$s$^{-3}$ (see Eq. [\ref{Dobro}]), which again
is lower than the observed heating.

\subsection{Alfv\'enic streams}

In paper I {\it Verma et al.} find that the observed temperature evolution
of the Alfv\'enic streams are in good agreement with the predictions of
Kolmogorov-like phenomenology if Kolmogorov's constant for MHD turbulence $%
C\ $is equal to 8.0. In paper I $C^{+}$ and $C^{-}$ were taken to be equal
which is inconsistent with the DIA calculations of {\it Verma and
Bhattacharjee} [1995b,c]. In the following discussion we modify the
dissipation rates of paper I using the theoretical results of {\it Verma and
Bhattacharjee} [1995b,c].

For large $\sigma _c$, since $\alpha _1$ is small, from Eq. (\ref{Kolm-diss}%
)
\begin{equation}
\epsilon _{turb}\sim \frac 12\frac{\sqrt{\alpha _1}}{C^{+}\sqrt{C^{-}}}%
\left[ E^{+}(k)\right] ^{3/2}k^{5/2}.
\end{equation}
If we take $\sigma _c=0.90$ ($\alpha _1=0.05$) and $r_A=2.0$, {\it Verma and
Bhattacharjee }[1995c] predict $C^{+}=1.54$ and $C^{-}=2.08$. With these
constants $\epsilon _{turb}$ predicted by the Kolmogorov-like phenomenology
is larger by a factor of $8^{3/2}/(C^{+}\sqrt{C^{-}})\sim 10$ as compared to
what was calculated in paper I, i.e., the predicted $\epsilon _{turb}$ with $%
C^{\pm }$ of DIA is $10^{-2}$ km$^2$s$^{-3}$. When we take $r_A=1.0$,
unfortunately, the maximum $\sigma _c$ we can get in DIA scheme is 0.5
(after $\sigma _c$ $=0.5\ $there is no consistent solution). This limitation
indicates that some of the assumptions of {\it Verma and Bhattacharjee }%
[1995b,c] are not fully consistent. However, from the general trend of $%
C^{\pm }$ we believe that the Kolmogorov-like MHD turbulence phenomenology
predicts higher dissipation rate as compared to what is observed in the
solar wind.

The above study appears to justify the conjecture of {\it Marsch }[1991] and
{\it Grappin et al.,} [1991] which states that in the solar wind the
Alfv\'enic streams have not reached steady-state, and the turbulence in
these streams is not yet fully developed. The nonlinear interactions may be
distributing energy among large and intermediate wavenumbers, with only a
small amount of energy flowing into the dissipation range. Therefore, the
real dissipation rate is lower than that predicted by the phenomenology,
which assumes that the system is in steady-state. Any modelling of
``approach towards steady-state turbulence'' will shed light to the
evolution of energy spectra and help us in correct estimation of turbulent
heating in the Alfv\'enic streams in the solar wind.

The simulational studies show that the system reaches steady-state after
around 1 eddy turnover time [{\it Verma}, 1994; {\it Verma et al.}, 1995d].
For solar wind it corresponds to the distance of approximately 2 AU $%
(1(l/u_{rms})V_{SW}=1*(0.1AU/20)*400=2AU)$. Hence, it is possible that the
Alfv\'enic streams have not reached steady-state during its transit in the
inner heliosphere, and the estimated turbulent heating by the turbulence
phenomenologies is higher than realistic heating in the solar wind. However,
it is hard to justify the above mentioned conjecture in the outer
heliosphere. Also note the nonAlfv\'enic streams appear to have reached
steady-state by the time they reach 1 AU. This could be due to the initial
conditions of the nonAlfv\'enic streams near the sun. We need to perform
careful analysis of the approach to steady-state and effects of cross
helicity in MHD turbulence to understand this behaviour.

Considering that the turbulent heating is not sufficient to heat the
nonAlfv\'enic streams to the observed temperature, in the following section
we consider the heating due to shocks and thermal diffusion.

\section{Heating due to Shocks and Thermal Diffusion}

{\it Whang et al. }[1990] studied the average entropy increase across a
shock for heliocentric distances of 1-15 AU. They showed that an average
entropy increase per shock, $\Delta S,$ is approximately $0.8\times 10^{-23}$
J/K/proton. The strong shocks were in the region of 1-5 AU. {\it Whang et
al.'}s [1990] found approximately 400 shocks in 1-15 AU region. Therefore,
the dissipation rate $\epsilon _{shock}=NT\Delta S/\Delta t$, where $N$ is
the number of shocks in 1-15 AU, $T$ is the temperature and $\Delta t$ is
the time taken to travel 1 AU to 15 AU, will be
\begin{equation}
\begin{array}{ll}
\epsilon _{shock} & \sim NT\Delta S/\Delta t \\
& \sim 400\times 10^5\times 0.8\times 10^{-23}/(14\times 1.5\times 10^8/500)
\mbox{J/proton} \\  & \sim 400\times 10^5\times 0.8\times 10^{-23}\times
10^{-3}/(m_p\times 14\times 1.5\times 10^8/500)
\mbox{km$^{2}$/s$^{3}$} \\  & \sim 10^{-3}\mbox{km$^{2}$/s$^{3}$}
\end{array}
\end{equation}
Here $m_p$ is the mass of a proton. From the above calculation we see that
shock heating is comparable to the turbulent heating, and they are one of
the major sources of heating in the outer heliosphere. However, note that
{\it Whang et al.'}s [1990] observational study of entropy increase in the
solar wind with the radial distance
\begin{equation}
S-S_E=3.81\times \log (r/r_E)10^{-23}\text{J/K/proton}
\end{equation}
where the subscript $E$ denotes the conditions at $r_0=$ 1\ AU, is the total
entropy increase due to all the heating sources present in the solar wind,
not just due to shocks, as inferred in the {\it Whang et al.}'s [1990] paper.

Regarding heat conduction, it is widely believed that the coupling between
electrons and protons is weak because the protons-electron collision
frequency is much smaller as compared to electron-electron or proton-proton
collision frequencies. Due to this reason, for the following discussion we
assume that the temperature of the protons in the solar wind are affected by
the heat flux of protons alone,  not by the heat flux of electrons. {\it %
Marsch and Richter }[1984] performed observational studies of the solar wind
at 0.95 AU and reported that the heat conduction by protons $q_i$ is speed
dependent and is approximately 10$^{-4}$ ergs/cm$^2$s (also refer to {\it %
Hundhausen }[1972], {\it Marsch et al. }[1983], and {\it Gazis }[1984]). At
0.95 AU, using particle density as 5/cc, the proton heating rate $\nabla
q_i/\rho $ due to this heat flux is 10$^{-4}$ km$^2$s$^{-3}$, which is
comparable to the heating rates due to turbulence. Hence, the dissipation
due to heat flux appears to be a likely candidate for the solar wind heating.

Now we will obtain theoretical estimates of the heating due to thermal
conduction using order of magnitude calculations and compare them with the
observational results. The source of heating due to thermal conduction,
according to Eq. (\ref{tempr}), is
\begin{equation}
H=-\frac{\nabla \cdot {\bf q}_i}\rho =\frac{\nabla \cdot (K_i\nabla T_i)}%
\rho \sim \frac{K_iT_i}{r^2\rho },
\end{equation}
where $K_i$ are the thermal conductivity of protons, $T_i$ are the
temperature of protons, and $r$ is the radial distance from the sun. Here we
take $r=1$ AU and $T_i=10^5K$. Note that the quantities $H$ decreases with
distance. The proton thermal conductivity $K_i$ $\sim \rho c_Vvl\sim \rho
c_V^i\nu $ [{\it Priest, }1982], where $c_V$ is the specific heat per unit
mass of protons, $v$ is the relevant velocity scale, $l$ is the
length-scale, and $\nu $ is the kinematic viscosity. Therefore,
\begin{equation}
\label{conduct}H=\frac{c_V\nu T_i}{r^2}.
\end{equation}
Note that $R=8.31$ J/K/mole, $c_V\ $for protons is 12.5 Joule/K/gm.

{\it Montgomery }[1983] estimated viscosity in the solar wind using the
classical transport theory of {\it Braginskii }[1965] and obtained typical $%
\nu $ $\sim $ 10$^{-6}$ km$^2$s$^{-1}$ [{\it Montgomery, }1983]. We assume
that the proton temperature are approximately $10^5$ K. With the above
viscosity we find that at 1 AU ($r=1$ AU), $H\sim 10^{-14}$ km$^2$s$^{-3}$.
This $H$ is negligible as compared to the turbulent dissipation rates and is
also inconsistent with the $H$ obtained from the observations [{\it Marsch
and Richter}, 1984; {\it Marsch et al.,} 1983; {\it Hundhausen, }1972].

However, since the solar wind plasma is turbulent, thermal conduction
coefficient should be determined by large-scale velocity and length scales [%
{\it Landau, }1987]. The basic idea is that in a turbulent plasma, the large
scale eddies carry heat flux from one region to the other. For this reason,
we must use turbulent eddy viscosity due to the large eddies in our
calculation for $K$. Therefore, $\nu \sim vl$, where $v$ $\sim 20$ km/s
(speed of large eddies) and $l$ $\sim 10^{12}$ cm (size of the large
eddies). This yields $H\sim 10^{-5}$ km$^2$s$^{-3}$. Since our calculations
involve only order of magnitude estimates, we can say that our estimates are
in general agreement with the observational results of {\it Marsch and
Richter }[1984], {\it Marsch et al.} [1983], and {\it Hundhausen }[1972].
Therefore, heating due to thermal conduction, corotating shocks, as well as
turbulence appear to be the heating sources of the solar wind in the outer
heliosphere.

{\it Gazis }[1984] estimated the thermal energy flux at 1 AU in the solar
wind and found it to be equal to $(2.5\pm 1.0)\times 10^{-2}$ ergs/cm$^2$s,
which corresponds to $\nabla \cdot {\bf q}/\rho \sim 2\times 10^{-2}$ km$^2$s%
$^{-3}$, that is approximately 10 times larger than the dissipation rate
estimated in paper\ I. However, note that in {\it Gazis'} formalism thermal
condition was considered as the only source of entropy increase (see Eq.
(24) of {\it Gazis }[1984]), therefore, the heating rate calculated by {\it %
Gazis }was an over-estimate. This is not surprising since {\it Gazis }[1984]
sets an upper limit on the total heat flux ${\bf q}$.

Regarding heating due to thermal conduction in the inner heliosphere, the
substitution of $T_i=10^6K$, $r=0.3$ AU, $v=20$ km/s, and $l=10^{12}$ cms in
Eq. (\ref{conduct}) yields $H_i\sim 10^{-3}$ km$^2$s$^{-3}$. This quantity
is comparable to turbulent heating rate in the solar wind. Thus, ion heating
due to thermal conduction is a potentially significant heating source in the
inner heliosphere and could provide significant part of the heating in the
inner heliosphere. Note that the corotating shocks are absent in the inner
heliosphere, therefore, shock heating is presumably not significant for
nonAlfv\'enic streams in the inner heliosphere. Of course, other sources,
e.g., stream-stream interactions, neutral ions etc. may also provide
significant heating in the inner heliosphere.

To summarize, turbulence provides only part of the dissipation in the solar
wind. Corotating shocks as well as thermal conduction appear to be other
heating sources in outer heliosphere and could provide rest of the heating.
In the inner heliosphere, heating due to thermal conduction appear to be
significant. In absence of corotating shocks in the inner heliosphere,
thermal conduction may provide significant part of the dissipation here. In
this paper we have performed only order of magnitude estimates of these
quantities, therefore, we cannot make any definite statements regarding
energy budget in the solar wind.

Turbulent heating is considered to be one of the potential mechanism which
could explain solar corona heating problem. In the following section we
estimate amount of turbulent heating in the solar corona using the above
mentioned MHD turbulence phenomenologies.

\section{Turbulent Heating in the Solar Corona}

The temperature of the sun decreases radially outwards to about 6000 K at
photosphere, then it rises abruptly upto $2-3\times 10^6\,$K at the corona.
Why the coronal temperature rises so steeply remains an open problem.
Various heating mechanisms have been proposed, turbulent heating being one
of them [{\it Hollweg,} 1984; {\it Browning and Priest}, 1984; {\it %
Browning, }1991 and references therein]. In the following discussion we
state {\it Hollweg}'s [1984] calculation of the dissipation rates in the
solar corona using the Kolmogorov-like phenomenology, and we calculate the
dissipation rates using Dobrowolny et al.'s generalized phenomenology. We
compare these estimates with the observational results. Here too we use
dimensional arguments similar to those used for solar wind. The power of
these arguments lies in the fact that we can estimate the dissipation rates
from the inertial range spectra or from the energy fed in at large-scales,
but without any knowledge of the dissipation mechanism which may be active
at small-scales.

Here we rephrase the {\it Hollweg's }arguments [1984] on the estimation of
turbulent heating in the corona in terms of $E^{\pm }(k).$ We assume that in
the corona $E^{+}(k)\sim E^{-}(k)\sim E(k)$. Therefore, the dissipation rate
according to the Kolmogorov-like phenomenology will be given by
\begin{equation}
\epsilon \sim C^{-3/2}(E(k)k)^{3/2}k.
\end{equation}
Taking the average speed of the large-scale eddies $v_k$ as 30 km/s, $%
E(k)k\sim v_k^2\sim 10^3$ km$^2$s$^{-2}$. The size of the largest eddies is
roughly $10^5$ km. Therefore, $\epsilon \sim 10$ km$^2\,$s$^{-3}$. {\it %
Hollweg }[1984] argues that the above dissipation rate is compatible with
the heating requirements of coronal active region loops.

In contrast, the dissipation rate in Dobrowolny et al.'s generalized
phenomenology [{\it Dobrowolny et al.}, 1980] will be
\begin{equation}
\epsilon =A^{-2}B_0^{-1}(E(k)k)^2k.
\end{equation}
With mean number density as $10^{15}$ particles/m$^3\,$and the magnetic
field of the order 10-100 mT, $B_0$ in velocity units is$\ 10^3-10^4$ km/s.
Taking $E(k)k\sim 10^3$ km$^2$s$^{-2}$ as before, we obtain $\epsilon \sim
10^{-2}-10^{-3}$ km$^2\,$s$^{-3}$. This dissipation rate is three to four
order of magnitude lower than that estimated by the Kolmogorov-like model.
Therefore, if turbulent heating is playing a major role in the coronal
heating, the Kolmogorov-like phenomenology must be at work in the solar
corona. This is surprising because one would expect Dobrowolny et al.'s
model to work due to the dominance of the mean magnetic field over the
fluctuations. In the solar wind too, even though the Parker field is
stronger than the fluctuations, the energy spectra follows roughly $%
k^{-5/3}\,$power law. Also the temperature evolution of the solar wind is
better explained by the Kolmogorov-like model rather than Dobrowolny et
al.'s phenomenology (see paper I). These observations indicate that
Kolmogorov-like phenomenology is applicable under conditions where it is not
expected to work.

The eddy turnover time-scale in the solar corona is $l/v_l\sim 10^5/30\sim
300$ sec. Since the plasma reaches steady-state in around 1 eddy-turnover
time [{\it Verma,} 1994; {\it Verma et al.}, 1995d], the question arises
whether it is possible for the plasma to reach steady-state and be able to
supply enough energy in the dissipation range. It is suggested that on the
solar surface and in the coronal loops plasma could stay for around 300
seconds, and therefore, turbulent heating could be one of the possible
candidate for coronal heating (see {\it G\'omez and Font\'an} [1988]; {\it %
Browning }[1991] and references therein). To resolve the above issues one
needs to resort to model calculations; for some of the recent calculations
refer to {\it G\'omez and Font\'an }[1988], {\it Heyvaert and Priest }%
[1992], {\it Zirker} [1993], {\it Vekstein et al. }[1993].

\section{Conclusions}

In this paper we have estimated the contributions of turbulent dissipation
in the heating of the solar wind using Kolmogorov's constants for MHD
turbulence calculated theoretically by {\it Verma and Bhattacharjee }%
[1995b,c]. We find that for the nonAlfv\'enic streams, turbulence
contributes only partly to the total heating in the solar wind. The
remaining part of the heating should be provided by corotating shocks,
thermal conduction, stream-stream interactions, interactions with the
neutral ions, and by other sources. The corotating shocks are present in the
outer heliosphere. As shown by {\it Whang et al. }[1990], they could be one
of the major heating sources in the outer heliosphere, at least in 1-15 AU
where they have been studied, and could provide a major fraction of the
heating. The order of magnitude calculations show that the heating due to
thermal conduction is significant both in the inner as well as outer
heliosphere. In absence of corotating shocks in the inner heliosphere,
thermal conduction probably plays an important role in the solar wind
heating and could possible provide significant fraction of the heating here.
Our results are based on order of magnitude estimates, therefore, we cannot
make definite statements. Models incorporating these results may help us in
making definite predictions about the contributions by these sources to the
solar wind heating.

For the Alfv\'enic streams, the Kolmogorov-like MHD turbulence phenomenology
with Kolmogorov's constants calculated by {\it Verma and Bhattacharjee }%
[1995b,c] predicts higher dissipation rate as compared to what is observed
in the wind. The resolution of the paradox could be that the Alfv\'enic
streams have not reached steady-state [{\it Marsch, }1991 and references
therein; {\it Grappin et al., }1991], and the energy is just being
distributed among various modes, with only a small amount of energy flowing
into the dissipation range and heating the plasma. Hence the dissipation
rates predicted by the Kolmogorov-like MHD turbulence phenomenology are
over-estimates. The studies on ``approach to steady-state'' will help us in
proper estimation of turbulent heating in the Alfv\'enic streams and in
understanding of other related problems.

Lastly, we have discussed the contributions of turbulent dissipation in the
coronal heating. {\it Hollweg }[1984] has calculated the heating rate using
Kolmogorov's fluid turbulence phenomenology and showed that the results are
consistent with the required energy flux in the coronal active loops. In
this paper we calculate the heating rates using {\it Dobrowolny et al.'}s
generalized model and show that the prediction from this model is four
orders of magnitude smaller than the required heating rates. This
calculation shows that if turbulent heating is major source of heating in
the corona, then the Kolmogorov-like phenomenology rather than {\it %
Dobrowolny et al.}'s phenomenology is valid in the corona. In the corona,
the mean magnetic field in velocity units is much larger than the
fluctuations, hence, if we assume that the heating in the solar corona is
due to turbulence, the this result is contrary to the assumptions of the
phenomenologies. However, the result is in agreement with the solar wind
results, where the Kolmogorov-like model works even though the Parker's
field is larger than the fluctuation. This is an important puzzle in MHD
turbulence, and future theoretical and numerical work will help in
clarifying this issue.

The author thanks D. A. Roberts, M. L. Goldstein, and J. K. Bhattacharjee
for discussion and comments.

\newpage\

\newpage

\begin{table} \centering
  \begin{tabular}{||c|c|c|c||} \hline
    $r_{A}$ & $\sigma_{c}$ & $C^{+}$ & $C^{-}$ \\ \hline
     0.5 & 0.0 & 4.04 & 4.04 \\
     1.0 & 0.0 & 3.38 & 3.38 \\
     2.0 & 0.0 & 2.58 & 2.58 \\
     5.0 & 0.0 & 1.92 & 1.92 \\
     1.0 & 0.43 & 0.54 & 53.80 \\
      1.5 & 0.67 & 1.44 & 6.47 \\
      2.0 & 0.90 & 1.54 & 2.08 \\ \hline
  \end{tabular}
  \caption{Kolmogorov's constants for MHD turbulence for various
values of Alfv\'{e}n ratio and normalized cross helicity [{\em Verma
and Bhattacharjee}, 1995b,c]}
\end{table}

\end{document}